\documentclass[5p]{elsarticle}
\usepackage{array}
\usepackage{bm}
\usepackage{bbm}
\usepackage{color}
\usepackage{graphicx}
\usepackage{amsmath,amssymb,amstext}
\numberwithin{equation}{section}   
\usepackage{multirow}
\usepackage{rotating}
\usepackage{float}
\usepackage{mathptmx}
\usepackage[numbers]{natbib}
\usepackage{soul,xcolor} 
\setul{-.6ex}{.3ex} 
\setstcolor{red}
\def\bra#1{\bigl\langle{ #1} \bigr|}
\def\ket#1{\bigl|{ #1} \bigr\rangle}

\def\ovlp#1#2{\bigl\langle{ #1}\big|{#2} \bigr\rangle}

\def\rvec {{\bf r}}
\def\pvec {{\bf p}}

\def\hvec {{\bf h}}
\def\qvec {{\bf q}}

\def\kvec {{\bf k}}

\def\sij#1#2{{\bm\sigma}_{#1}\cdot{\bm\sigma}_{#2}}

\def\KF{k_{\rm F}}
\def\tF{t_{\rm F}}
\def\SF{S_{\rm F}}
\def\a0{a_0}
\def\I{{\rm i}}

\def\ie{{\em i.e.\/}\ }

\def\1{\mathbbm{1}}

\begin{document}
\begin{frontmatter}
  
\title{The nature of short-ranged correlations in nuclear systems}

\author{E.~Krotscheck and J. Wang}

\address{$\dagger$Department of Physics, University at Buffalo, SUNY
Buffalo NY 14260}
\address{$^\ddagger$Institut f\"ur Theoretische Physik, Johannes
Kepler Universit\"at, A 4040 Linz, Austria}

\begin{abstract}
  We apply ideas of the parquet-diagram and optimized
  Fermi-hypernetted chain methods to determine the short-range
  structure of the pair wave function in neutron matter and compare
  these with Bethe-Goldstone results and those of low-order
  variational calculations.  It is shown that the induced interaction,
  describing the exchange of density, spin, and tensor fluctuations,
  has a profound influence on the short-ranged structure of the pair
  wave function and, hence, on effective interactions in neutron
  matter.
\end{abstract}

\ead{eckhardk@buffalo.edu}

\end{frontmatter}
\newpage

\section{Introduction}
\label{sec:intro}
It is generally understood that the nature of the wave function in a
many-body system is, at short interparticle distances, determined by a
Schr\"odinger-like equation.  The prime example for this is the
time-honored Bethe-Goldstone equation
\cite{BRU55a,BRU55,BetheGoldstone57,Goldstone57,BruecknerLesHouches}
which is basically a 2-body Schr\"odinger equation modified by the
Pauli principle. Many-body effects are mostly included through the
single-particle spectrum. The literature on the subject matter is
vast, we cite here only early works \cite{BetheBrandowPetschek}
and review articles \cite{MahauxReview,MahauxSartor}.

Similar in spirit is the low-order version of Jastrow-Feenberg theory.
The method begins with a variational {\em ansatz\/} for the many-body
wave function. One assumes a semi-realistic interaction of the form
\begin{equation}
\hat v (i,j) = \sum_{\alpha=1}^n v_\alpha(r_{ij})\,
        \hat O_\alpha(i,j),
\label{eq:vo}
\end{equation}
where $r_{ij}=|\rvec_i-\rvec_j|$ is the distance between particles $i$
and $j$, and the $O_\alpha(i,j)$ are operators acting on the spin,
isospin, and possibly angular momentum variables of the individual
particles.  Semi-realistic model potentials use the six
base operators are
\begin{align}
O_1(i,j;\hat\rvec_{ij})
        &\equiv \1,\qquad
O_3(i,j;\hat\rvec_{ij})
        \equiv ({\bm\sigma}_i \cdot {\bm\sigma}_j)\,,
\nonumber\\
O_5(i,j;\hat\rvec_{ij})
&\equiv S_{ij}(\hat\rvec_{ij})
      \equiv3({\bm\sigma}_i\cdot \hat\rvec_{ij})
      ({\bm\sigma}_j\cdot \hat\rvec_{ij})-{\bm\sigma}_i \cdot {\bm\sigma}_j\,,
      \nonumber\\
      O_{2n}(i,j;\hat\rvec_{ij}) &= O_{2n-1}(i,j;\hat\rvec_{ij})
      {\bm\tau}_1\cdot{\bm\tau}_2\,.
\label{eq:operator_v6}
\end{align}
where $\hat\rvec_{ij} = \rvec_{ij}/r_{ij}$. An appropriate variational wave
function is in this case
\begin{equation}
        \Psi_0^{{\rm SOP}}
        = {\cal S} \Bigl[ \prod^N_{i,j=1 \atop i<j} \hat f (i,j)\Bigr] \Phi_0\,,
\label{eq:f_prodwave}
\end{equation}
where ${\cal S}$ stands for symmetrization, and
\begin{equation}
  \hat f(i,j) = \sum_{\alpha=1}^n f_\alpha(r_{ij})\,
  O_\alpha(i,j)\,.\label{eq:falpha}
\end{equation}
The need to symmetrize the operator product in \eqref{eq:f_prodwave}
causes considerable complications and no summation method has been
found that comes anywhere close to the diagrammatic richness of the
hypernetted-chain (HNC) summation method for bosons \cite{LGB} and
fermions \cite{Mistig,Fantoni74,MistigNP,Fantoni} that has been
achieved for the case of purely central correlations. The components
$f_\alpha(r_{ij})$ of the correlation operator $\hat f (i,j)$ are
therefore often determined by minimization of the two-body
approximation of $E_0$, subject to a healing constraint
\cite{ScottMozowski}. The method is then referred to as ``low order
constrained variational method
(LOCV)''\cite{PandharipandeBethe,Johnreview}.  The energy expectation
value is calculated either in low order, or by partial diagram
summations like the ``single-operator-chain (SOC)'' method
\cite{Pand76,IndianSpins}.  Both the Bethe-Goldstone equation and the
LOCV determination of the correlations imply only a minimal inclusion
of many-body effects; the fact that they led to rather different
answers has caused significant discussions
\cite{ClarkTriesteSummary,DickhoffMBT18}.

The situation is much simpler for the case of central correlations.
Most importantly, the summation method defines a hierarchy of
approximations which permit an unconstrained variational determination
of the correlations by minimization of the energy expectation value
$E_0$ \cite{FeenbergBook},
\begin{equation}
\frac{\delta E_0}{\delta f}(\rvec_1,\rvec_2) = 0\,.
\label{eq:euler}
\end{equation}
For fermions, some care must be taken in the treatment of exchange
diagrams to permit an unconstrained variation \cite{EKVar}.

A further important insight was that the hypernetted chain summation
method together with the unconstrained optimization \eqref{eq:euler}
corresponds, for bosons, to a self-consistent summation of both the
ring and the ladder diagrams of perturbation theory
\cite{parquet1,parquet2,parquet3}, in other words the Euler equation
\eqref{eq:euler} contains both the Bethe-Goldstone equation and the
RPA equation. This was already observed by Sim, Buchler, and Woo
\cite{Woo70}. The analogy is less systematic for fermions and implies
more approximations, but it was similarly proven for the most
important classes, namely rings, ladders, and self-energy corrections
\cite{fullbcs}.

Similar statements on the connection between diagrammatic expansions
for the symmetrized operator product wave function
\eqref{eq:f_prodwave} and Feynman diagrams are not available. One of
the reasons is, of course, the lack of a summation procedure that is
as complete as the HNC summations for state-independent correlation
and would permit an unconstrained optimization of the components
$f_\alpha(r)$ of the correlation operator \eqref{eq:falpha}. The
second, more subtle reason is that there is evidence that the
commutator terms introduced by the symmetrization procedure actually
correspond to non-parquet diagrams \cite{SpinTwist}.

In view of these complications, Smith and Jackson \cite{SmithSpin}
started from the idea of parquet-diagram summations and implemented
the procedure for a fictive system of bosonic nucleons interacting via
a $v_6$ interaction. We have recently followed up on that idea
\cite{v3eos} and extended it to fermions. In generalizing the parquet
equations to fermions, we have used the diagrammatic ideas of the
parquet-diagram summations, and taken, when necessary or appropriate,
approximations suggested by variational wave functions. Most
importantly, the form \eqref{eq:f_prodwave} implies that all two-body
functions depend only on the distance.  Since the Fermi sea defines a
preferred reference frame, such local functions can be obtained only
by specific Fermi-sea averages, we will mention these where
appropriate.

\section{A brief survey of FHNC-EL and parquet equations}
As stated above, the FHNC-EL and parquet diagram summation implies the
self-consistent summation of both ring-- and ladder diagrams. To be
more specific, the ring diagrams are derived from a random-phase
approximation (RPA) equation for the response function
 \begin{align}
    \chi(q,\omega) &=
    \frac{\chi_0(q,\omega)} {1-\hat V_{\rm
        p-h}(q)\chi_0(q,\omega)}\nonumber\\
 S(q) &= -\int_0^\infty \frac{d\hbar\omega}{\pi} \Im m \chi(q,\omega),
\label{eq:SRPA}
 \end{align}
 in terms of a local ``particle-hole'' interaction $\hat V_{\rm
   p-h}(q)$. In the case of state-dependent interactions, $\hat V_{\rm
   p-h}(q)$ is a linear combination of local functions $\tilde
 V^{(\alpha)}_{\rm p-h}(q)$ and the base operators
 \eqref{eq:operator_v6}. In \eqref{eq:SRPA} it is more convenient to
 represent $\hat V_{\rm p-h}(q)$ in the basis of the operators $\1$,
 $\hat L_{12} =
 \frac{1}{3}\left(\sij{1}{2}+S_{12}(\hat\rvec_{12})\right)$ and $\hat
 T_{12} = \frac{1}{3}\left(\sij{1}{2}-2S_{12}(\hat\rvec_{12})\right)$.
 As usual we define the dimensionless Fourier transform by including a
 density factor $\rho$, \ie $ \tilde f(q) = \rho\int d^3r e^{\I
   \qvec\cdot\rvec}f(r)$.

 The second relevant relationship is the Bethe-Goldstone equation for
 the pair wave function
\begin{align}
  \bra{\kvec,\kvec'}\psi\ket{\hvec,\hvec'}
  &=   \ovlp{\kvec,\kvec'}{\hvec,\hvec'}\label{eq:fullpsi2}\\
  &- \bar n(k)\bar n(k')\frac{\bra{\kvec,\kvec'}v\psi
    \ket{\hvec,\hvec'}}{e(k) + e(k') -e(h)-e(h')}
\nonumber\,.
\end{align}
Above, the $e(k)$ are the single-particle energies. We use the
convention that $\hvec, \hvec', \hvec_i$ are occupied (``hole'')
states, $\pvec,\pvec',\pvec_i$ are unoccupied (``particle'') states
whereas $\kvec$, $\qvec$ have no restriction.

In making the connection to FHNC-EL, we must define a pair
wave function that is a function of the relative coordinate
or momentum, \ie it has the feature that
\[
\bra{\kvec,\kvec'}\psi\ket{\hvec,\hvec'} =
\frac{1}{N}\tilde\psi\left(\left|\kvec-\hvec\right|\right)
\equiv \frac{1}{N}\tilde\psi(q)\,.
  \]
Similarly, for local interactions, we should have
\[
\bra{\kvec,\kvec'}v\psi\ket{\hvec,\hvec'} = \frac{1}{N}
\left[v(r)\psi(r)\right]^{\cal F}(q)\,.
\]
where $\left[\ldots\right]^{\cal F}$ stands for Fourier transform. To ensure this, the energy denominator in Eq. \eqref{eq:fullpsi2} must
somehow be approximated by a function of momentum transfer $q
=\left|\kvec-\hvec\right|$. Bethe and Goldstone
\cite{BetheGoldstone57} set the center of mass momentum to zero, see
also chapter 11 in Ref. \citenum{FetterWalecka}.  Closer to the spirit
of variational theory is to approximate the energy denominator by its
Fermi-sea average
\begin{equation}
  \tF(q) = \frac{\sum_\hvec (1-n(|\hvec+\qvec|))(t(|\hvec+\qvec|) -t(h))}{\sum_\hvec
    (1-n(|\hvec+\qvec|))} = \frac{\hbar^2 q^2}{2m\SF(q)}\,.
\end{equation}
Eq. \eqref{eq:fullpsi2} can then be rewritten in the form
\begin{equation}
  \left[\frac{\hbar^2}{m}\nabla^2 + v(r)\right]\psi(r)
  = \frac{\rho}{\nu}\int d^3r' \ell^2(|\rvec-\rvec'|\KF)v(r')\psi(r')
  \label{eq:BGlocal}
\end{equation}
where $\ell(x) = \frac{3}{x}j_1(x)$, and $\nu$ is the degree of
degeneracy of the single-particle states, and $\SF(q)=1
-\frac{\rho}{\nu} \int d^3r e^{\I\qvec\cdot\rvec}\ell^2(r\KF)$ is
the static structure function of non-interacting fermions. The version
of Eq. \eqref{eq:fullpsi2} for zero center of mass momentum differs
from Eq. \eqref{eq:BGlocal} only by the fact that
$\ell^2(|\rvec-\rvec'|\KF)$ is replaced by $\ell(|\rvec-\rvec'|\KF)$.

Summing the parquet diagrams supplements, among others, the bare
interaction $\hat v(r)$ by an induced interaction $\hat w_I(r)$ being
defined as the set of particle-hole reducible diagrams. Assuming a
particle-hole interaction $\hat V_{\rm p-h}(q)$ of the operator of the
form \eqref{eq:vo}, the sum of these diagrams is {\em a priori} an
energy-dependent quantity
\begin{equation}
  \hat w_I(q,\omega) =  \frac{\hat V_{\rm p-h}^2(q)\chi_0(q,\omega)}
  {1-\hat V_{\rm p-h}(q)\chi_0(q,\omega)}\,.
\end{equation}
The connection between parquet theory and HNC-EL is then made by
defining \cite{parquet1,parquet2} an energy independent effective
interaction $\hat w(q)$ as follows:
Calculate the static structure function
\begin{align}
  S(q) &= - \int_0^\infty \frac{d\hbar\omega}{\pi}\,
  \Im m\frac{\chi_0(q,\omega)}{1- \chi_0(q,\omega)\hat V_{\rm p-h}(q)}
  \nonumber\\
  & = - \int_0^\infty \frac{d\hbar\omega}{\pi}\,
  \Im m\left[\chi_0(q,\omega) + \chi_0^2(q,\omega)\hat w_I(q,\omega)\right]\,.
\end{align}
Now define an {\em energy independent interaction\/} $\hat
w_I(q,\bar\omega(q))$ by demanding that it gives the same static
structure function,
\begin{equation}
  S(q)\equiv -\int_0^\infty \frac{d\hbar\omega}{\pi}\,
  \Im m\left[\chi_0(q,\omega) + \chi_0^2(q,\omega)\hat w_I(q,\bar\omega(q))\right]\,.
\label{eq:Scond}
\end{equation}

This energy independent interaction $\hat w_I(q) \equiv \hat
w_I(q,\bar\omega(q))$ is then taken as a correction to the interaction
in the Bethe-Goldstone equation. For state-dependent interactions it
is again understood that $\hat w_I(q,\omega)$ is a linear combination
of local functions and operators of the form \eqref{eq:vo}.  For
solving the Bethe-Goldstone equation, it is convenient to write both
the interaction and the induced interaction as a linear combination of
the spin-singlet projector $\hat P_s = (\1 - \sij{1}{2})/4$, and the
spin-triplet projectors $\hat P_{t+} = (3\1 + \sij{1}{2} +
S_{12}(\hat\rvec))/6$ and $\hat P_{t-} = (3\1 + \sij{1}{2} -2
S_{12}(\hat\rvec))/12$.

We also mention briefly the connection to the FHNC-EL summation method.
Diagrammatically we can identify the pair wave function $\psi(r)$ with
the direct correlation function $\Gamma_{\!\rm dd}(r)$
\begin{equation}
  \psi(r) = \sqrt{1+\Gamma_{\!\rm dd}(r)}\,.\label{eq:psiGdd}
\end{equation}
The equation determining the short-ranged structure of $\psi(r)$ or
$\sqrt{1+\Gamma_{\!\rm dd}(r)}$ is slightly different from
Eq. \eqref{eq:BGlocal}, see Eq. (3.33) of Ref. \citenum{fullbcs}; we
found, however, in our numerical applications that the numerical
solutions are very close. We shall, therefore, not elaborate on
this issue any further.

\section{Short-ranged structure}

Let us now go through a step-by-step analysis of the influence of the
induced interaction and the consequences for other wuantities for a
specific example We have chosen the Reid $v_6$ interaction
\cite{Reid68} in the parametrization of Day \cite{Day81} for neutron
matter at the relatively low density $\KF = 1\,\mathrm{fm}^{-1}$. We
have carried out a sequence of calculations
\begin{enumerate}
\item{} Simply set $\hat w_I(r)$ zero. That correspond to the Bethe-Goldstone
  equation and, in a sense, also to LOCV.
\item{} Use the state-dependent theory, Eq. \eqref{eq:Scond} \cite{v3eos}.
\item{} Use the FHNC-EL (or parquet) version for purely central
  correlations as described, for example, in Refs. \citenum{fullbcs}
  or \citenum{ectpaper}. This implies that only the central component
  the interaction operator \eqref{eq:vo} is kept.
\item{} Take the spin-singlet component of the interaction operator in
  the Bethe-Goldstone equation but use the induced interaction from
  the state-independent calculation.
\end{enumerate}

\begin{figure}[h]
  \includegraphics[width=0.6\columnwidth,angle=-90]{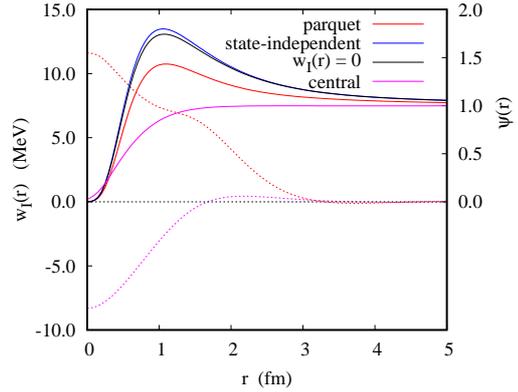}
  \caption{(Color online) The figure shows, for the Reid $v_6$
    interaction at $\KF = 1.0\,\mathrm{fm}^{-1}$, the induced
    interaction for the case of fully state-dependent parquet
    calculation (red dashed line, left axis) and the corresponding
    pair wave function $\psi(r)$ (red solid line, right axis). Also
    shown are the induced interaction and the pair wave function
    \eqref{eq:psiGdd} coming from a state-independent parquet
    calculation using the central component of $v_6$ only (magenta
    dashed and solid lines). The blue line shows the pair wave
    function when the induced interaction is taken from the
    state-independent calculation, but the spin-singlet interaction is
    used instead of the central interaction.  Note that this version
    does not lead to self-consistent solutions of the parquet
    equation.  Also shown is the pair wave function if the induced
    interaction is omitted (black solid line).}\label{fig:psiplot}
\end{figure}
\section{Results}

Turning to the key message of our paper, we emphasize that the
essential additional quantity provided by parquet theory or FHNC-EL is
the induced interaction $\hat w_I(q)$. FHNC-EL or parquet offers the
prescriptions \eqref{eq:Scond} for calculating this
quantity. Fig. \ref{fig:psiplot} shows the the singlet component if
the induced interactions $w_I(r)$ coming from a state-dependent and
state-independent parquet calculation, along with the resulting pair
wave functions \eqref{eq:psiGdd}.  Since the central part of the Reid
potential is less attractive than the singlet projection, the pair
wave function for the state-independent case has no pronounced nearest
neighbor peak.

To isolate the relevance of the induced interaction, we have also
carried out a calculation taking the spin-singlet projection in the
Bethe-Goldstone equation, but the induced interaction from the
state-independent calculation. The result is shown in blue in Fig.
\ref{fig:psiplot}. This calculation is somewhat inconsistent at the
state-independent parquet-level because there the correlations should
be determined by the central part of the of the interaction operator.
The resulting pair wave function is close to the one obtained by
simply omitting the induced interaction. However, taking the
spin-singlet component of the interaction in a full parquet
calculation does not lead to self-consistent solutions of the parquet
equations even at very low densities. Hence, it is very important to
include the induced interaction to guarantee the stability of the
system.

We have also tried the original version of Bethe and Goldstone,
setting the center of mass momentum to zero. The results are very
similar to those shown in Fig. \ref{fig:psiplot}, in particular the
key message of our paper on the importance of the induced interaction
remains unchanged.

Pair wave functions are auxiliary quantities, effective interactions
are more directly relevant for dynamic properties and pairing
phenomena. The essential input is always the particle-hole
interaction, which also determines the induced interaction
$w_I(r)$. It is therefore important to both verify the validity of our
analysis and to determine the importnace of the pair wave function for
these interactions.  In terms of the quantities introduced above, the
simplest version, dubbed FHNC-EL//0, has the form
\begin{align}
  V_{p-h}^{(\alpha)}(r) &= \left[1+\Gamma_{\rm dd}^{(\alpha)}(r)\right]v_\alpha(r)
+ t^{(\alpha)}_{\rm CW}(r)
+ \Gamma_{\rm dd}^{(\alpha)}(r)w_I^{(\alpha)}(r)\,.\label{eq:Vphdef}\nonumber\\
  t^{(\alpha)}_{\rm CW}(r)&\equiv \frac{\hbar^2}{m}\left|\nabla\sqrt{1+\Gamma_{\rm dd}^{(\alpha)}(r)}\right|^2
  \end{align}
where $\alpha$ stands for the spin-singlet and the two spin-triplet
projections, and $t^{(\alpha)}_{\rm CW}(r)$ is the 2-body part of the
``Clark-Westhaus'' kinetic energy \cite{Johnreview}. In the
state-independent approximation, all $\Gamma_{\rm dd}^{(\alpha)}(r)$
are the same. We display the form \eqref{eq:Vphdef} for the sake of
discussion of the essential effects, exchange corrections are
important and have been included in our calculation as described in
Refs. \citenum{fullbcs} and \citenum{v3eos}.

The expression \eqref{eq:Vphdef} shows nicely the physical effects that
contribute to the particle-hole interaction. These were described by
Aldrich and Pines \cite{Aldrich,ALP78}:
\begin{enumerate}
\item{} Local screening of the short-ranged repulsion of the bare
  interaction.  This is described by the factor $ 1+\Gamma_{\!\rm
    dd}^{(\alpha)}(r)$, which goes to zero at short distances.
\item{} The cost in kinetic energy to bend down the wave function.
  This leads to some repulsion and a ``swelling'' of the core.  The
  effect is described by the kinetic energy term $t_{\rm
    CW}^{(\alpha)}(r)$.
\item{} An enhancement of the attraction due to the presence of other
  particles. Again, the factor $1+\Gamma_{\rm dd}^{(\alpha)}(r)$
  describes this effect, note that an attractive interaction generates
  a high nearest neighbor peak in the pair wave function and, hence,
  enhances the attractive part of the interaction.
\item{} The last term $\Gamma_{\rm dd}^{(\alpha)}(r)w_I^{(\alpha)}(r)$
  describes the modificaction of the interactions through the exchange
  of density or spin fluctuations.
\end{enumerate}

Returning to Fig. \ref{fig:psiplot}, one might wonder why a relatively
small correction to the bare interaction can have a rather profound
effect.  After all, the well-depth of the spin-singlet Reid
interaction is of the order of 100 MeV.  One of the reasons for the
sensitive dependence of the pair wave function on the induced
interaction is that the bare singlet interaction has almost a bound
state, the S-wave scattering length is $-18.7\ $fm
\cite{PhysRevLett.83.3788}, in other words the 2-body interaction is
close to developing a bound state and small changes of the interaction
can cause large changes in the pair wave function.

\begin{figure}[ht]
  \centerline{\includegraphics[width=0.6\columnwidth,angle=-90]{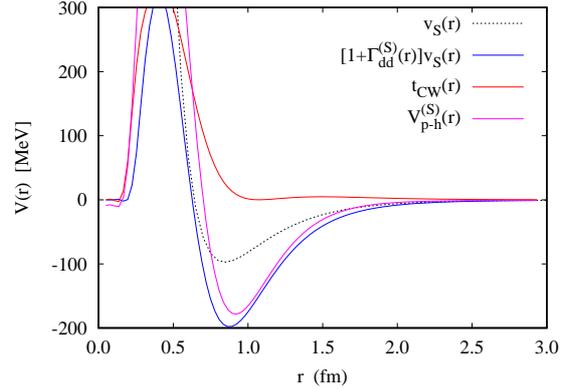}}
  \caption{(Color online) The figure shows, for the Reid $v_6$ interaction
    at $\KF = 1.0\,\mathrm{fm}^{-1}$ in the spin-singlet
    channel the bare interaction $v_S(r)$ (black dashed line), the components
    $\left[1+\Gamma_{\rm dd}^{(S)}(r)\right]v_S(r)$ and $t_{\rm CW}^{(S)}(r)$
    (blue and red solid line, respectively) and the full $V_{\rm p-h}^{(S)}(r)$
    (magenta line).}\label{fig:Vphplot}
  \end{figure}

The most important issue for our discussion is the enhancement of the
effective interaction due to the enhanced nearest neighbor peak as
shown in Fig. \ref{fig:psiplot}. Such an enhancement can, for example,
have significant effects on pairing phenomena in neutron matter which
depend sensitively on the interaction strength. To see this effect, we
show in Fig. \eqref{fig:Vphplot} the bare singlet interaction and the
components of the particle hole interaction \ref{eq:Vphdef}. The most
pronounced effect is evidently the enhancement of the interaction by
the peak in the pair wave function. The kinetic energy term $t_{\rm
  CW}^{(\alpha)}(r)$ basically causes the ``swelling'' of the core,
but is not strong enough to compensate for the enhancement of the
attraction.

The long-wavelength limits of the components of the particle-hole
interactions can be related to Landau's Fermi-liquid parameters,

\begin{equation}
  \tilde V_{\rm p-h}^{(c)}(0+) = mc_{\rm F}^{*2}F_0^s,\qquad
  \tilde V_{\rm p-h}^{(\sigma)}(0+) = mc_{\rm F}^{*2}F_0^a,\qquad
\,,
\label{eq:FLPfromVph}
\end{equation}
where $c_{\rm F}^* = \sqrt{\frac{\hbar^2\KF^2}{3mm^*}}$ is the speed
of sound of the non-interacting Fermi gas with the effective mass
$m^*$, and $F_0^{s,a}$ are Landau's Fermi liquid parameters.

The two Landau parameters can also be obtained by derivatives of the equation
of state as a function of density and spin-polarization, for example
we have for the incompressibility
\begin{equation}
  mc^2 = \frac{d}{d\rho}\rho^2 \frac{d}{d\rho}\frac{E}{N}
  = mc_{\rm F}^{*2}(1+F_0^S)\,.
\label{eq:mcfromeos}
\end{equation}
Predictions for the
Fermi-liquid parameters derived from hydrodynamic derivatives and from
effective interactions agree only in an exact theory
\cite{EKVar,parquet5}; good agreement is typically reached only at
very low densities \cite{fullbcs}.

\begin{figure}[h]
  \centerline{\includegraphics[width=0.6\columnwidth,angle=-90]{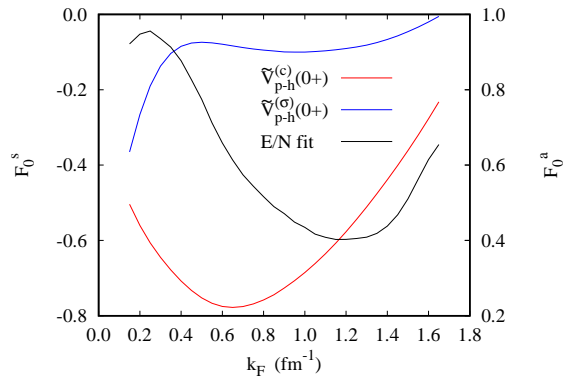}}
    \caption{(Color online) The figure shows the Fermi-liquid
      parameters $F_0^s$ (red line, left scale) and $F_0^a$ (blue
      line, right scale) as obtained from the particle hole
      interaction, Eqs. \eqref{eq:FLPfromVph} for the Reid $v_6$
      potential as a function of density. Also shown is $F_0^s$ as
      obtained from the equation of state via Eq. \eqref{eq:mcfromeos}
      (black line, left scale).}
    \label{fig:F0aF0aplot}
\end{figure}

Fig. \ref{fig:F0aF0aplot} shows the calculated Fermi-Liquid parameters
$F_0^s$ and $F_0^a$ for the Reid $v_6$ model potential as obtained
from Eqs. \eqref{eq:FLPfromVph} and \eqref{eq:mcfromeos}. Note that
the results for $F_0^s$ differ slightly from those of Ref. \citenum{v3eos}
due to improved numerics.

A relationship similar to Eq. \eqref{eq:mcfromeos} can be derived for
$F_0^a$ by calculating the equation of state of a partially
spin-polarized system. Such a calculation goes beyond the scope of
this paper, we note however that our results indicate that $F_0^a$ is
positive and of the order of 1 which agrees with several previous
calculations. \cite{PLB2491990,EPL171992,PhysRevC.87.014338}.

\begin{figure}[h]
  \centerline{\includegraphics[width=0.45\columnwidth,angle=-90]{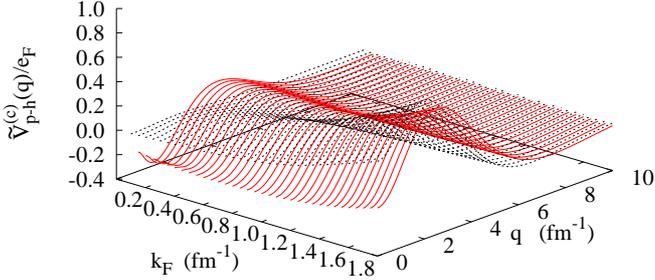}}
  \caption{(Color online) The figure shows, for the Reid $v_6$
    interaction, the central channel of the particle-hole interaction
    obtained by the full parquet calculation (red, solid lines) and
    the state-independent approximation (black, dashed lines). The
    interaction strength is given in units of the Fermi energy of the
    non-interacting system,
    $\hbar^2\KF^2/2m$.}\label{fig:re_Vphef_c_3d}
\end{figure}

\begin{figure}[h]
  \centerline{\includegraphics[width=0.45\columnwidth,angle=-90]{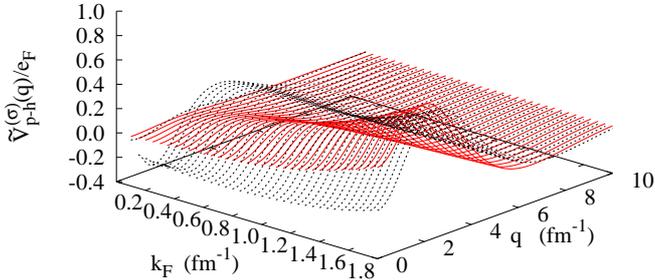}}
  \caption{(Color online) Same as Fig. \ref{fig:re_Vphef_c_3d} for the
    spin-channel interaction $\tilde V_{\rm p-h}^{(\sigma)}(q)$.}
  \label{fig:re_Vphef_sigma_3d}
\end{figure}

Returning to the full particle-hole interaction, we show in
Figs. \ref{fig:re_Vphef_c_3d} and \ref{fig:re_Vphef_sigma_3d} a
comparison between $\tilde V_{\rm p-h}^{(\alpha)}(q)$ for the central and the
spin-channel for both the state-dependent and the state-independent
theory. Recall that in the state-independent approximation, all
$\Gamma_{\rm dd}^{(\alpha)}(r) \equiv \Gamma_{\rm dd}(r)$ and
$w_I^{(\alpha)}(r) \equiv w_I(r)$ are the same, and we get simply
$V_{\rm p-h}^{(\sigma)}(r) = \left[1+\Gamma_{\rm
    dd}(r)\right]v_\sigma(r)$.  Evidently the agreement between
results from state-dependent and state-independent calculations is at
most qualitative.

\section{Summary}

We have in this paper highlighted the importance of the exchange of
density and spin fluctuations for the short-ranged structure of
correlations in neutron matter and their immediate relevance for
effective interactions. We have chosen the example of neutron matter
because this is -- next to electrons and perhaps cold gases away from
the uniform limit -- one of the simplest realistic many-body
systems. The reason for this simplicity is that neutron natter is, as
opposed to liquid $^3$He and nuclear matter, not self-bound.  A
self-bound Fermi system has necessarily at least two spinodal points
below saturation density which have the immediate consequence that the
equation of state is a non-analytic function of the density. These
complications do not exist in neutron matter and we can focus on the
problem at hand, namely the importance and the treatment of
operator-dependent correlations.

Immediate applications are foreseeable, among others, for pairing
properties of neutron matter which has been discussed for decades, for
a collection of recent review articles, see Ref. \citenum{ECTSI}.
Similarly important is the response of neutron matter which has been
discussed over the years, \cite{Wam93,Benhar2009,Lovato2013}.

A completely open issue, which must be resolved before these phenomena
are examined, is the importance of non-parquet diagrams corresponding
to the commutator diagrams in the wave function
\eqref{eq:f_prodwave}. There is evidence \cite{SpinTwist} that these
diagrams can be very important whenever the core size of the
interaction is very different in different interaction channels. This
is the case for both the Reid and the Argonne interactions.

\section*{Acknowledgment}
 This work was supported, in part, by the the College of Arts and
 Sciences of the University at Buffalo, SUNY. Encouragement for this
 work was derived from a workshop on {\em Nuclear Many-Body Theories:
   Beyond the mean field approaches\/} at the Asia Pacific Center for
 Theoretical Physics in Pohang, South Korea, in July 2019. We thank
 J. W. Clark for numerous comments and suggestions on this
 manuscript. One of us (JW) thanks the Austrian Marshall Plan
 Foundation for support during the summer 2018 and Robert Zillich for
 discussions.


\begin{thebibliography}{45}
\expandafter\ifx\csname natexlab\endcsname\relax\def\natexlab#1{#1}\fi
\providecommand{\url}[1]{\texttt{#1}}
\providecommand{\href}[2]{#2}
\providecommand{\path}[1]{#1}
\providecommand{\DOIprefix}{doi:}
\providecommand{\ArXivprefix}{arXiv:}
\providecommand{\URLprefix}{URL: }
\providecommand{\Pubmedprefix}{pmid:}
\providecommand{\doi}[1]{\href{http://dx.doi.org/#1}{\path{#1}}}
\providecommand{\Pubmed}[1]{\href{pmid:#1}{\path{#1}}}
\providecommand{\bibinfo}[2]{#2}
\ifx\xfnm\relax \def\xfnm[#1]{\unskip,\space#1}\fi
\bibitem[{Brueckner(1955{\natexlab{a}})}]{BRU55a}
\bibinfo{author}{K.~A. Brueckner}, \bibinfo{journal}{Phys. Rev.}
  \bibinfo{volume}{97} (\bibinfo{year}{1955}{\natexlab{a}})
  \bibinfo{pages}{1353--1365}.
\bibitem[{Brueckner(1955{\natexlab{b}})}]{BRU55}
\bibinfo{author}{K.~A. Brueckner}, \bibinfo{journal}{Phys. Rev.}
  \bibinfo{volume}{100} (\bibinfo{year}{1955}{\natexlab{b}})
  \bibinfo{pages}{36--45}.
\bibitem[{Bethe and Goldstone(1957)}]{BetheGoldstone57}
\bibinfo{author}{H.~A. Bethe}, \bibinfo{author}{J.~Goldstone},
  \bibinfo{journal}{Proc. R. Soc. London, Ser. A} \bibinfo{volume}{238}
  (\bibinfo{year}{1957}) \bibinfo{pages}{551--567}.
\bibitem[{Goldstone(1957)}]{Goldstone57}
\bibinfo{author}{J.~Goldstone}, \bibinfo{journal}{Proc. R. Soc. London, Ser. A}
  \bibinfo{volume}{239} (\bibinfo{year}{1957}) \bibinfo{pages}{267--279}.
\bibitem[{Brueckner(1959)}]{BruecknerLesHouches}
\bibinfo{author}{K.~A. Brueckner}, in: \bibinfo{editor}{C.~DeWitt},
  \bibinfo{editor}{P.~Nozi{\`e}res} (Eds.), \bibinfo{booktitle}{Lecture Notes
  of the 1957 Les Houches Summer School}, \bibinfo{publisher}{Dunod},
  \bibinfo{year}{1959}, pp. \bibinfo{pages}{47--241}.
\bibitem[{Bethe et~al.(1963)Bethe, Brandow, and
  Petschek}]{BetheBrandowPetschek}
\bibinfo{author}{H.~A. Bethe}, \bibinfo{author}{B.~H. Brandow},
  \bibinfo{author}{A.~G. Petschek}, \bibinfo{journal}{Phys. Rev.}
  \bibinfo{volume}{129} (\bibinfo{year}{1963}) \bibinfo{pages}{225--264}.
\bibitem[{Jeukenne et~al.(1976)Jeukenne, Lejeune, and Mahaux}]{MahauxReview}
\bibinfo{author}{J.~P. Jeukenne}, \bibinfo{author}{A.~Lejeune},
  \bibinfo{author}{C.~Mahaux}, \bibinfo{journal}{Physics Reports}
  \bibinfo{volume}{25} (\bibinfo{year}{1976}) \bibinfo{pages}{83}.
\bibitem[{Mahaux and Sartor(1992)}]{MahauxSartor}
\bibinfo{author}{C.~Mahaux}, \bibinfo{author}{R.~Sartor},
  \bibinfo{journal}{Physics Reports} \bibinfo{volume}{211}
  (\bibinfo{year}{1992}) \bibinfo{pages}{53--211}.
\bibitem[{Scott and Moszkowski(1962)}]{ScottMozowski}
\bibinfo{author}{B.~L. Scott}, \bibinfo{author}{S.~A. Moszkowski},
  \bibinfo{journal}{Nucl. Phys.} \bibinfo{volume}{29} (\bibinfo{year}{1962})
  \bibinfo{pages}{665--671}.
\bibitem[{Pandharipande and Bethe(1973)}]{PandharipandeBethe}
\bibinfo{author}{V.~R. Pandharipande}, \bibinfo{author}{H.~A. Bethe},
  \bibinfo{journal}{Phys. Rev. C} \bibinfo{volume}{7} (\bibinfo{year}{1973})
  \bibinfo{pages}{1312--1328}.
\bibitem[{Clark(1979)}]{Johnreview}
\bibinfo{author}{J.~W. Clark}, in: \bibinfo{editor}{D.~H. Wilkinson} (Ed.),
  \bibinfo{booktitle}{Progress in Particle and Nuclear Physics},
  volume~\bibinfo{volume}{2}, \bibinfo{publisher}{Pergamon Press Ltd.},
  \bibinfo{address}{Oxford}, \bibinfo{year}{1979}, pp.
  \bibinfo{pages}{89--199}.
\bibitem[{Pandharipande and Wiringa(1976)}]{Pand76}
\bibinfo{author}{V.~R. Pandharipande}, \bibinfo{author}{R.~B. Wiringa},
  \bibinfo{journal}{Nucl. Phys. A} \bibinfo{volume}{266} (\bibinfo{year}{1976})
  \bibinfo{pages}{269--316}.
\bibitem[{Pandharipande and Wiringa(1979)}]{IndianSpins}
\bibinfo{author}{V.~R. Pandharipande}, \bibinfo{author}{R.~B. Wiringa},
  \bibinfo{journal}{Rev. Mod. Phys.} \bibinfo{volume}{51}
  (\bibinfo{year}{1979}) \bibinfo{pages}{821--859}.
\bibitem[{Clark(1979)}]{ClarkTriesteSummary}
\bibinfo{author}{J.~W. Clark}, \bibinfo{journal}{Nucl. Phys. A}
  \bibinfo{volume}{328} (\bibinfo{year}{1979}) \bibinfo{pages}{587--595}.
\bibitem[{Dickhoff(2016)}]{DickhoffMBT18}
\bibinfo{author}{W.~H. Dickhoff}, \bibinfo{journal}{Journal of Physics:
  Conference Series} \bibinfo{volume}{702} (\bibinfo{year}{2016})
  \bibinfo{pages}{012013}.
\bibitem[{van Leeuwen et~al.(1959)van Leeuwen, Groeneveld, and Boer}]{LGB}
\bibinfo{author}{J.~M.~J. van Leeuwen}, \bibinfo{author}{J.~Groeneveld},
  \bibinfo{author}{J.~D. Boer}, \bibinfo{journal}{Physica} \bibinfo{volume}{25}
  (\bibinfo{year}{1959}) \bibinfo{pages}{792--808}.
\bibitem[{Krotscheck and Ristig(1974)}]{Mistig}
\bibinfo{author}{E.~Krotscheck}, \bibinfo{author}{M.~L. Ristig},
  \bibinfo{journal}{Phys. Lett. A} \bibinfo{volume}{48} (\bibinfo{year}{1974})
  \bibinfo{pages}{17--18}.
\bibitem[{Fantoni and Rosati(1974)}]{Fantoni74}
\bibinfo{author}{S.~Fantoni}, \bibinfo{author}{S.~Rosati},
  \bibinfo{journal}{Lett. Nuovo Cimento} \bibinfo{volume}{10}
  (\bibinfo{year}{1974}) \bibinfo{pages}{545--551}.
\bibitem[{Krotscheck and Ristig(1975)}]{MistigNP}
\bibinfo{author}{E.~Krotscheck}, \bibinfo{author}{M.~L. Ristig},
  \bibinfo{journal}{Nucl. Phys. A} \bibinfo{volume}{242} (\bibinfo{year}{1975})
  \bibinfo{pages}{389--405}.
\bibitem[{Fantoni and Rosati(1975)}]{Fantoni}
\bibinfo{author}{S.~Fantoni}, \bibinfo{author}{S.~Rosati},
  \bibinfo{journal}{Nuovo Cimento} \bibinfo{volume}{25A} (\bibinfo{year}{1975})
  \bibinfo{pages}{593--615}.
\bibitem[{Feenberg(1969)}]{FeenbergBook}
\bibinfo{author}{E.~Feenberg}, \bibinfo{title}{Theory of {Q}uantum Fluids},
  \bibinfo{publisher}{Academic}, \bibinfo{address}{New York},
  \bibinfo{year}{1969}.
\bibitem[{Krotscheck(1977)}]{EKVar}
\bibinfo{author}{E.~Krotscheck}, \bibinfo{journal}{Phys. Rev. A}
  \bibinfo{volume}{15} (\bibinfo{year}{1977}) \bibinfo{pages}{397--407}.
\bibitem[{Jackson et~al.(1982)Jackson, Lande, and Smith}]{parquet1}
\bibinfo{author}{A.~D. Jackson}, \bibinfo{author}{A.~Lande},
  \bibinfo{author}{R.~A. Smith}, \bibinfo{journal}{Physics Reports}
  \bibinfo{volume}{86} (\bibinfo{year}{1982}) \bibinfo{pages}{55--111}.
\bibitem[{Jackson et~al.(1985)Jackson, Lande, and Smith}]{parquet2}
\bibinfo{author}{A.~D. Jackson}, \bibinfo{author}{A.~Lande},
  \bibinfo{author}{R.~A. Smith}, \bibinfo{journal}{Phys. Rev. Lett.}
  \bibinfo{volume}{54} (\bibinfo{year}{1985}) \bibinfo{pages}{1469--1471}.
\bibitem[{Krotscheck et~al.(1986)Krotscheck, Smith, and Jackson}]{parquet3}
\bibinfo{author}{E.~Krotscheck}, \bibinfo{author}{R.~A. Smith},
  \bibinfo{author}{A.~D. Jackson}, \bibinfo{journal}{Phys. Rev. A}
  \bibinfo{volume}{33} (\bibinfo{year}{1986}) \bibinfo{pages}{3535--3536}.
\bibitem[{Sim et~al.(1970)Sim, Woo, and Buchler}]{Woo70}
\bibinfo{author}{H.~K. Sim}, \bibinfo{author}{C.-W. Woo},
  \bibinfo{author}{J.~R. Buchler}, \bibinfo{journal}{Phys. Rev. A}
  \bibinfo{volume}{2} (\bibinfo{year}{1970}) \bibinfo{pages}{2024--2037}.
\bibitem[{Fan and Krotscheck(2019)}]{fullbcs}
\bibinfo{author}{H.-H. Fan}, \bibinfo{author}{E.~Krotscheck},
  \bibinfo{journal}{Physics Reports} \bibinfo{volume}{823}
  (\bibinfo{year}{2019}) \bibinfo{pages}{1--59}.
\bibitem[{Krotscheck(1988)}]{SpinTwist}
\bibinfo{author}{E.~Krotscheck}, \bibinfo{journal}{Nucl. Phys. A}
  \bibinfo{volume}{482} (\bibinfo{year}{1988}) \bibinfo{pages}{617--652}.
\bibitem[{Smith and Jackson(1988)}]{SmithSpin}
\bibinfo{author}{R.~A. Smith}, \bibinfo{author}{A.~D. Jackson},
  \bibinfo{journal}{Nucl. Phys. A} \bibinfo{volume}{476} (\bibinfo{year}{1988})
  \bibinfo{pages}{448--470}.
\bibitem[{Krotscheck and Wang(2020)}]{v3eos}
\bibinfo{author}{E.~Krotscheck}, \bibinfo{author}{J.~Wang},
  \bibinfo{title}{Variational and parquet-diagram calculations for neutron
  matter. {I.} {S}tructure and energetics}, \bibinfo{year}{2020}.
  \bibinfo{note}{{P}hys. Rev. C (in press)}.
\bibitem[{Fetter and Walecka(1971)}]{FetterWalecka}
\bibinfo{author}{A.~L. Fetter}, \bibinfo{author}{J.~D. Walecka},
  \bibinfo{title}{{Q}uantum Theory of Many-Particle Systems},
  \bibinfo{publisher}{McGraw-Hill}, \bibinfo{address}{New York},
  \bibinfo{year}{1971}.
\bibitem[{{Reid, Jr.}(1968)}]{Reid68}
\bibinfo{author}{R.~V. {Reid, Jr.}}, \bibinfo{journal}{Ann. Phys. (NY)}
  \bibinfo{volume}{50} (\bibinfo{year}{1968}) \bibinfo{pages}{411--448}.
\bibitem[{Day(1981)}]{Day81}
\bibinfo{author}{B.~D. Day}, \bibinfo{journal}{Phys. Rev. C}
  \bibinfo{volume}{24} (\bibinfo{year}{1981}) \bibinfo{pages}{1203--1271}.
\bibitem[{Fan et~al.(2017)Fan, Krotscheck, and Clark}]{ectpaper}
\bibinfo{author}{H.-H. Fan}, \bibinfo{author}{E.~Krotscheck},
  \bibinfo{author}{J.~W. Clark}, \bibinfo{journal}{J. Low Temp. Phys.}
  \bibinfo{volume}{189} (\bibinfo{year}{2017}) \bibinfo{pages}{470--494}.
\bibitem[{Gonz\'alez~Trotter et~al.(1999)Gonz\'alez~Trotter, Salinas, Chen,
  Crowell, Gl\"ockle, Howell, Roper, Schmidt, \ifmmode~\check{S}\else
  \v{S}\fi{}laus, Tang, Tornow, Walter, Wita\l{}a, and
  Zhou}]{PhysRevLett.83.3788}
\bibinfo{author}{D.~E. Gonz\'alez~Trotter}, \bibinfo{author}{F.~Salinas},
  \bibinfo{author}{Q.~Chen}, \bibinfo{author}{A.~S. Crowell},
  \bibinfo{author}{W.~Gl\"ockle}, \bibinfo{author}{C.~R. Howell},
  \bibinfo{author}{C.~D. Roper}, \bibinfo{author}{D.~Schmidt},
  \bibinfo{author}{I.~\ifmmode~\check{S}\else \v{S}\fi{}laus},
  \bibinfo{author}{H.~Tang}, \bibinfo{author}{W.~Tornow},
  \bibinfo{author}{R.~L. Walter}, \bibinfo{author}{H.~Wita\l{}a},
  \bibinfo{author}{Z.~Zhou}, \bibinfo{journal}{Phys. Rev. Lett.}
  \bibinfo{volume}{83} (\bibinfo{year}{1999}) \bibinfo{pages}{3788--3791}.
\bibitem[{Aldrich and Pines(1976)}]{Aldrich}
\bibinfo{author}{C.~H. Aldrich}, \bibinfo{author}{D.~Pines},
  \bibinfo{journal}{J. Low Temp. Phys.} \bibinfo{volume}{25}
  (\bibinfo{year}{1976}) \bibinfo{pages}{677--690}.
\bibitem[{{Aldrich III} and Pines(1978)}]{ALP78}
\bibinfo{author}{C.~H. {Aldrich III}}, \bibinfo{author}{D.~Pines},
  \bibinfo{journal}{J. Low Temp. Phys.} \bibinfo{volume}{31}
  (\bibinfo{year}{1978}) \bibinfo{pages}{689--715}.
\bibitem[{Jackson and Smith(1987)}]{parquet5}
\bibinfo{author}{A.~D. Jackson}, \bibinfo{author}{R.~A. Smith},
  \bibinfo{journal}{Phys. Rev. A} \bibinfo{volume}{36} (\bibinfo{year}{1987})
  \bibinfo{pages}{2517--2518}.
\bibitem[{Niembro et~al.(1990)Niembro, Marcos, Quelle, and
  Navarro}]{PLB2491990}
\bibinfo{author}{R.~Niembro}, \bibinfo{author}{S.~Marcos},
  \bibinfo{author}{M.~L. Quelle}, \bibinfo{author}{J.~Navarro},
  \bibinfo{journal}{Phys. Lett. B} \bibinfo{volume}{249} (\bibinfo{year}{1990})
  \bibinfo{pages}{373--376}.
\bibitem[{Cugnon et~al.(1992)Cugnon, Deneye, and Lejeune}]{EPL171992}
\bibinfo{author}{J.~Cugnon}, \bibinfo{author}{P.~Deneye},
  \bibinfo{author}{A.~Lejeune}, \bibinfo{journal}{Europhysics Letters}
  \bibinfo{volume}{17} (\bibinfo{year}{1992}) \bibinfo{pages}{129--132}.
\bibitem[{Holt et~al.(2013)Holt, Kaiser, and Weise}]{PhysRevC.87.014338}
\bibinfo{author}{J.~W. Holt}, \bibinfo{author}{N.~Kaiser},
  \bibinfo{author}{W.~Weise}, \bibinfo{journal}{Phys. Rev. C}
  \bibinfo{volume}{87} (\bibinfo{year}{2013}) \bibinfo{pages}{014338}.
\bibitem[{Krotscheck(2017)}]{ECTSI}
\bibinfo{editor}{E.~Krotscheck} (Ed.), \bibinfo{title}{Pairing and Condensation
  in {F}ermionic Systems}, volume \bibinfo{volume}{189} of
  \textit{\bibinfo{series}{J. Low Temp. Phys.}}, \bibinfo{publisher}{Springer},
  \bibinfo{address}{New York}, \bibinfo{year}{2017}.
\bibitem[{Wambach et~al.(1993)Wambach, Ainsworth, and Pines}]{Wam93}
\bibinfo{author}{J.~Wambach}, \bibinfo{author}{T.~Ainsworth},
  \bibinfo{author}{D.~Pines}, \bibinfo{journal}{Nucl. Phys. A}
  \bibinfo{volume}{555} (\bibinfo{year}{1993}) \bibinfo{pages}{128--150}.
\bibitem[{Benhar and Farina(2009)}]{Benhar2009}
\bibinfo{author}{O.~Benhar}, \bibinfo{author}{N.~Farina},
  \bibinfo{journal}{Phys. Lett. B} \bibinfo{volume}{680} (\bibinfo{year}{2009})
  \bibinfo{pages}{305--309}.
\bibitem[{Lovato et~al.(2013)Lovato, Losa, and Benhar}]{Lovato2013}
\bibinfo{author}{A.~Lovato}, \bibinfo{author}{C.~Losa},
  \bibinfo{author}{O.~Benhar}, \bibinfo{journal}{Nucl. Phys. A}
  \bibinfo{volume}{901} (\bibinfo{year}{2013}) \bibinfo{pages}{22--50}.

\end{thebibliography}
\end{document}